\documentclass[aps,preprint,groupedaddress,nofootinbib]{revtex4}

\usepackage[colorlinks=true, a4paper=true, pdfstartview=FitV,
linkcolor=blue, citecolor=blue, urlcolor=blue]{hyperref}

\usepackage{amsmath}
\usepackage{amssymb}
\usepackage{graphicx}
\usepackage{hhline}
\usepackage{mwe}
\usepackage{amsbsy}
\usepackage{textcomp}
\usepackage{commath}
\usepackage{subfig}
\usepackage{slashed}
\usepackage{natbib}
\usepackage{xcolor}

\usepackage{stackengine}
\stackMath

\makeatletter
\usepackage{babel}

\begin{document}
\immediate\write16{<<WARNING: LINEDRAW macros work with emTeX-dvivers
                    and other drivers supporting emTeX \special's
                    (dviscr, dvihplj, dvidot, dvips, dviwin, etc.) >>}

\title{Resonance windows in rapidly collapsing false vacuum bubbles}

\author{J. G. F. Campos}
\email{joao.gfcampos@upe.br}
\affiliation{Física de Materiais, Universidade de Pernambuco, Rua Benfica, 455, Recife - PE - 50720-001, Brazil}

\author{A. Mohammadi}
\email[]{azadeh.mohammadi@ufpe.br}
\affiliation{Departamento de Física, Universidade Federal de Pernambuco, Av. Prof. Moraes Rego, 1235, Recife - PE - 50670-901, Brazil}

\author{Fabiano C. Simas}
\email[]{fc.simas@ufma.br}
\affiliation{Departamento de Física, Universidade Federal do Maranhão, Av. dos Portugueses, 1966, São Luís - MA - 65080-805, Brazil.}

\begin{abstract}
A collapsing bubble of true vacuum does not necessarily disappear: it
can bounce and expand indefinitely, triggering the decay of the false
vacuum. We study this process in two and three spatial dimensions $D$ by
perturbing spherical sphalerons, the critical bubbles of a scalar field
theory, along their unstable mode. Weakly perturbed bubbles collapse
into long-lived oscillons, while sufficiently strong perturbations lead
to single-bounce vacuum decay. Between these regimes, multi-bounce
resonance windows accumulate geometrically, unlike in kink-antikink
collisions. Using an energy-exchange map, we relate this behavior to
the logarithmic divergence of the time spent near the sphaleron, which
fixes the scaling ratio in terms of the sphaleron's linear spectrum. Initially exciting the bound states shifts
the threshold for vacuum decay. For quadrupolar excitations, the shift
is even in the amplitude for $D=2$, while for $D=3$
prolate and oblate bubbles have different thresholds.
\end{abstract}

\maketitle

\section{Introduction}
\label{intro}

Kinks, oscillons \cite{gleiser1994pseudostable, copeland1995oscillons},
and sphalerons are among the simplest and most relevant solutions in
nonlinear field theory \cite{manton2004topological, vachaspati2007kinks}.
In $(1+1)$ dimensions they are closely related: a sphaleron can be
viewed as a kink-antikink pair, and if the pair is brought closer
together, their mutual attraction turns the configuration into an
oscillon \cite{manton2023simplest}.
The dynamics and scattering of kinks \cite{dorey2017boundary,
moradi2017multi, gani2018scattering, izquierdo2019kink,
christov2019long, adam2019phi4, christov2021kink, dorey2021resonance,
campos2021wobbling, campos2023kink, bazeia2023kink, dorey2023collisions},
oscillons \cite{graham2006unnatural, fodor2008small, hahne2020scattering,
dorey2024oscillons, simas2025vacuum}, and sphalerons
\cite{lima2021scattering, alonso2023semi, oles2023sphaleron,
navarro2024impact, anco2025long} in $(1+1)$ dimensions have been
studied extensively. Soliton dynamics in two and three spatial
dimensions, is also under active
investigation, with recent progress in vortex \cite{bazeia2024abelian,
krusch2024scattering, alonso2024spectral, alonso2025dynamics,
bachmaier2026resonance, martinez2026oscillons, alonso2026bps} and
monopole \cite{bachmaier2025simulations} collisions.

One of the simplest higher-dimensional systems with vacuum decay is a
real scalar field with a false vacuum. A spherical bubble of true
vacuum tends to expand because of the vacuum energy difference, while
the tension of its wall tends to make it contract. When the two effects
balance, one obtains the spherical sphaleron: an unstable static
solution, which is the critical bubble of thermal false-vacuum decay
\cite{linde1983decay}. A small perturbation determines whether the
bubble expands, collapses, or forms a long-lived oscillon. Oscillons,
domain walls, and bubbles in higher-dimensional scalar theories have
been studied numerically, for instance, in
Refs.~\cite{kasuya2003balls, hindmarsh2006numerical,
saffin2007oscillons, hindmarsh2008oscillons, easther2009new,
gleiser2011generation, kawasaki2015adiabatic}. In particular, Honda and Choptuik found that the lifetime of oscillons formed by collapsing $\phi^4$ bubbles in three spatial dimensions
diverges logarithmically as the initial bubble radius approaches
critical values, where the evolution approaches a periodic solution
with a single unstable mode \cite{honda2002fine}.

Resonance windows can arise in non-integrable models through energy exchange
between collective motion and localized modes, quasinormal modes, delocalized
modes, or external degrees of freedom~\cite{campbell1983resonance,
dorey2011kink, dorey2018resonant, adam2019spectral, adam2021sphalerons,
alonso2021scattering, adam2022spectral, blaschke2024amplitude}. In a recent
work~\cite{campos2026collision}, resonance windows and false-to-true vacuum
decay were found in collisions of oscillons generated from perturbed
sphalerons and from the small-amplitude expansion proposed
in~\cite{fodor2008small}. Moreover, sphaleron collisions in $(1+1)$
dimensions can produce false-vacuum bubbles bounded by a kink-antikink pair,
which repeatedly collapse and re-expand before decaying into an
oscillon~\cite{martinezsanchez2026false}. These results raise two questions:
whether the decay of a single perturbed sphaleron in higher dimensions shows
similar resonant behavior, and under which conditions a collapsing bubble can
trigger false-vacuum decay.

Unlike in $(1+1)$ dimensions, rapidly collapsing bubbles cannot be
obtained by Lorentz boosting a static solution. Instead, we use the
unstable mode of the spherical sphaleron to generate collapsing bubbles
in a controlled way, and study how the outcome depends on the
perturbation amplitude and on the excitation of the sphaleron's bound
states. We find long-lived oscillons, single-bounce vacuum decay, and,
in between, multi-bounce resonance windows. Unlike in kink-antikink
collisions, these windows accumulate geometrically at the threshold,
with a ratio fixed by the growth rate of the unstable mode and the
frequency of the radial bound state.

The paper is organized as follows.
In Sec.~\ref{sec:mod}, we introduce a scalar field model with a false and a true vacuum and numerically construct its spherical sphalerons. In Sec.~\ref{stability}, we analyze their linear stability and determine the spectrum of unstable and bound modes. In Sec.~\ref{sec:map}, we formulate an energy-exchange map and use it to predict the geometric accumulation of resonance windows in collapsing bubbles. In Sec.~\ref{sec:num}, we investigate the nonlinear decay of sphalerons perturbed along the unstable mode, compare the numerical results with the theoretical
predictions, and examine the effects of radial and quadrupolar bound-state excitations. In Sec.~\ref{sec:conclusion}, we summarize our main results and present our conclusions. In Appendix~\ref{App:K-AK coll}, we show that the same map reproduces the power-law accumulation of resonance
windows in kink-antikink collisions.

\section{The model}
\label{sec:mod}

\begin{figure}
    \centering
    \includegraphics[width=0.55\linewidth]{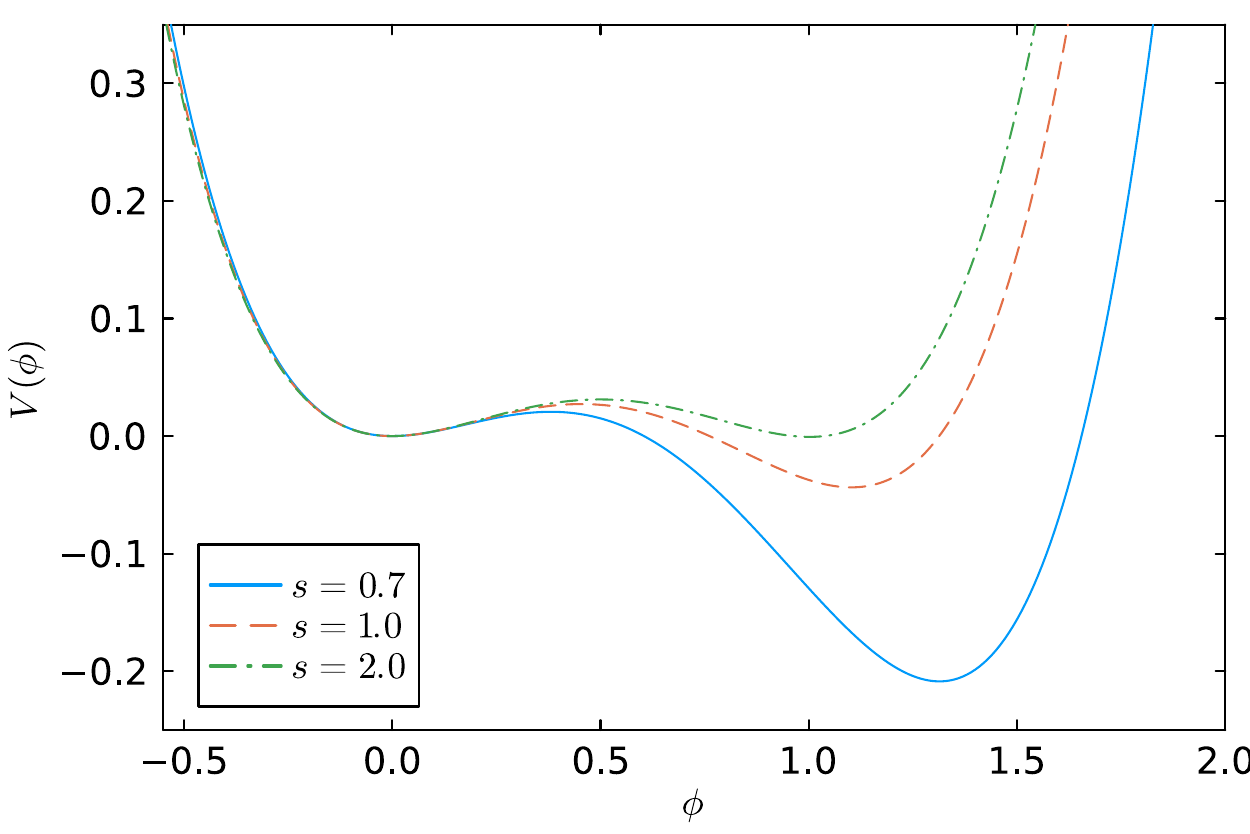}
    \caption{Potential $V(\phi;s)$ for several values of $s$. The false
    vacuum $\phi=0$ and the true vacuum $\phi_+$ are separated by the
    barrier top $\phi_-$.}
    \label{fig:potential}
\end{figure}

We consider a real scalar field $\phi(\mathbf{x},t)$ in $D+1$
dimensions, with Lagrangian density
\begin{equation}
\mathcal{L}=\frac{1}{2}\partial_\mu\phi\,\partial^\mu\phi-V(\phi;s)
\end{equation}
and potential~\cite{avelar2008new}
\begin{equation}
    V(\phi;s)
    =\frac{\phi^2}{2}\left(\phi-\tanh s\right)\left(\phi-\coth s\right)
    =\frac{\phi^2}{2}\left(1-2A\phi+\phi^2\right),
    \label{pot1}
\end{equation}
where $A=\coth 2s>1$ and $s>0$. The equation of motion is
\begin{equation}
    \phi_{tt}-\nabla^2\phi=-V'(\phi)
    =-\phi\left(1-3A\phi+2\phi^2\right).
    \label{eq:eom}
\end{equation}
Besides the false vacuum $\phi=0$, with $V(0)=0$ and unit mass,
$V''(0)=1$, the potential has two extrema,
\begin{equation}
    \phi_{\pm}=\frac{3A\pm\sqrt{9A^2-8}}{4}.
    \label{eq:extrema}
\end{equation}
The local maximum $\phi_-$ is the top of the barrier, and the global
minimum $\phi_+$, with $V(\phi_+)<0$, is the true vacuum
(Fig.~\ref{fig:potential}). The parameter $s$ controls the asymmetry:
as $s$ decreases, the true vacuum becomes deeper and moves to larger
$\phi$. For $s\to\infty$, $A\to1$ and
$V\to\frac{1}{2}\phi^2(\phi-1)^2$, the degenerate $\phi^4$ potential.

\begin{figure}
    \centering
    \includegraphics[width=0.48\linewidth]{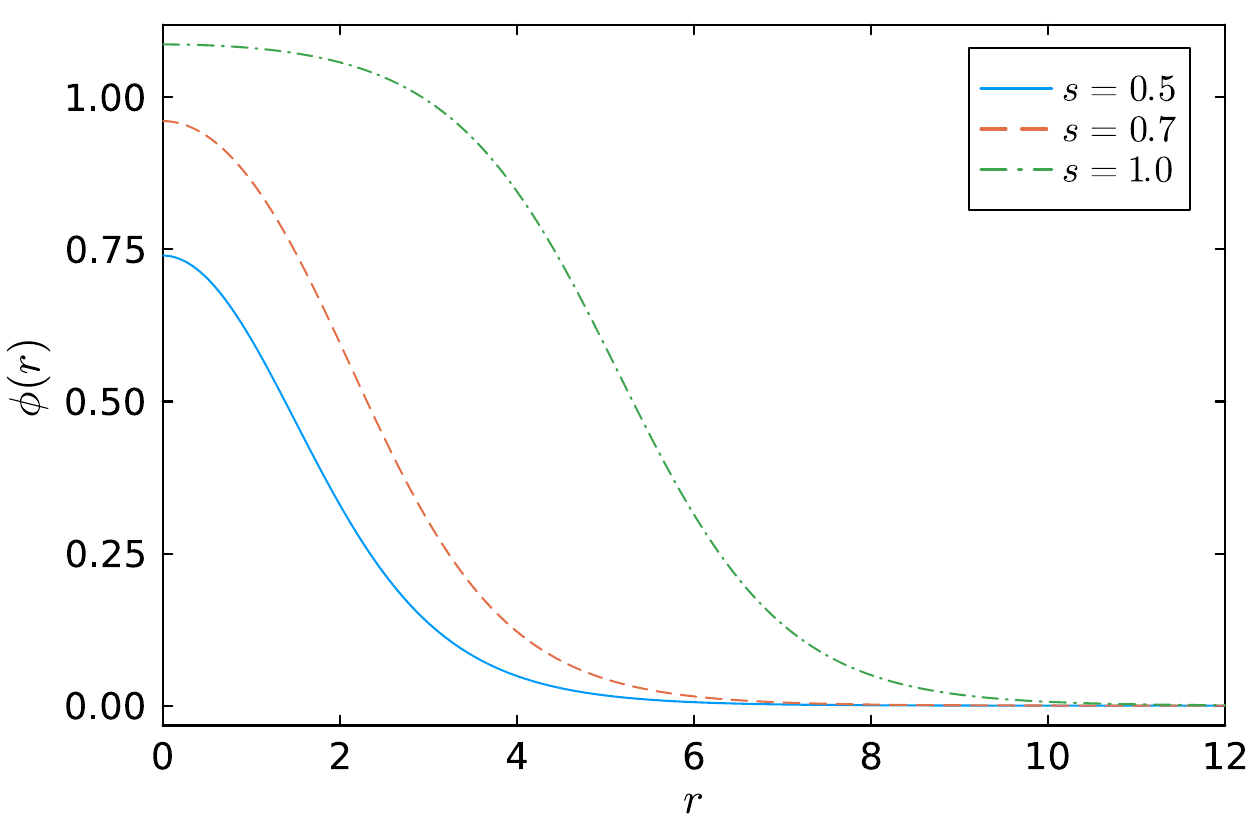}
    \includegraphics[width=0.48\linewidth]{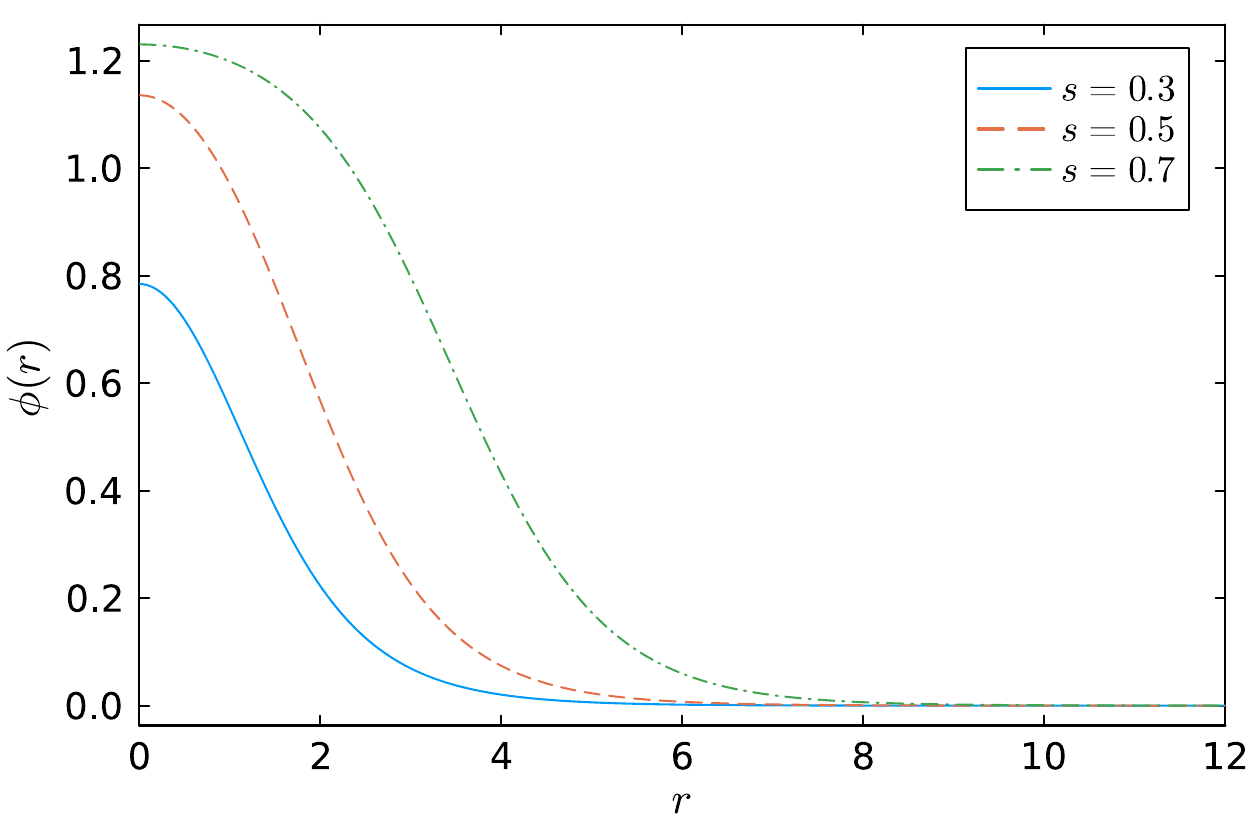}
    \caption{Sphaleron profiles $\phi_S(r)$ for several values of $s$ in
    $D=2$ (left) and $D=3$ (right).}
    \label{fig:sphaleron}
\end{figure}

We look for static, spherically symmetric solutions $\phi_S(r)$ of
\begin{equation}
    \phi_{rr}+\frac{D-1}{r}\phi_r=V'(\phi),
    \qquad
    \phi_r(0)=0,
    \qquad
    \phi(r\to\infty)=0.
    \label{eq-static}
\end{equation}
These are bubbles of true vacuum embedded in the false vacuum, in
which the pressure of the true vacuum balances the tension of the
wall. The balance is unstable: they are the critical bubbles of
false-vacuum decay, which are called sphalerons. Unlike kinks,
they are not topological, since the field approaches the false vacuum in all directions.
Figure~\ref{fig:sphaleron} shows the profiles for $D=2$ and $D=3$. 
They grow in amplitude and width with $s$ and with $D$. As $s\to\infty$, the vacuum energy difference
$\varepsilon=-V(\phi_+)$ vanishes while the wall tension $\sigma$
remains finite, so the bubble radius $R\simeq(D-1)\sigma/\varepsilon$
grows without bound. In this case the sphalerons approach thin-wall
bubbles~\cite{coleman1977fate,coleman1977erratum}.
For the parameters studied
here, however, $\phi_S(0)<\phi_+$. Therefore, the bubble core never reaches the true vacuum, and the sphalerons are far from the thin-wall regime.

\section{Linear stability analysis}
\label{stability}

We expand small perturbations of the sphaleron in hyperspherical
harmonics on $S^{D-1}$,
\begin{equation}
    \phi(r,\Omega,t)=\phi_S(r)
    +\epsilon\,\eta_{n,l}(r)\,Y_{l,\mathbf{m}}(\Omega)\,e^{-i\omega t},
    \qquad |\epsilon|\ll1,
    \label{eq:perturbation}
\end{equation}
where $-\Delta_{S^{D-1}}Y_{l,\mathbf{m}}=l(l+D-2)Y_{l,\mathbf{m}}$,
with $l=0,1,2,\dots$ and $\mathbf{m}$ the remaining angular quantum
numbers. For $D=2$ these are $\cos l\theta$ and $\sin l\theta$, and
for $D=3$ the spherical harmonics $Y_l^m(\theta,\varphi)$. To linear
order, Eq.~\eqref{eq:eom} gives
\begin{equation}
    -\eta_{rr}-\frac{D-1}{r}\eta_r
    +\left[\frac{l(l+D-2)}{r^2}+U(r)\right]\eta=\omega^2\eta,
    \qquad U(r)=V''\bigl(\phi_S(r)\bigr),
    \label{eq:stability}
\end{equation}
with $\eta\sim r^l$ as $r\to0$ and $\eta\to0$ as $r\to\infty$. Since
$U(r)\to V''(0)=1$ at infinity, the continuum starts at $\omega^2=1$.
Discrete eigenvalues with $0<\omega^2<1$ are stable bound states,
while $\omega^2<0$ indicates an instability with growth rate
$\Gamma=\sqrt{-\omega^2}$. For each $l$, the eigenvalues are ordered
as $\omega_{0,l}^2<\omega_{1,l}^2<\cdots$, where $n$ is the number of
radial nodes of $\eta_{n,l}$.

Negative eigenvalues can occur only for $l=0$. Translations,
$\delta\phi=\epsilon_i\partial_i\phi_S=\epsilon_i(x_i/r)\phi_S'(r)$,
give a zero mode in the $l=1$ sector with radial profile
$\eta_{0,1}\propto\phi_S'$. Since $\phi_S'$ has no nodes for $r>0$,
it is the lowest $l=1$ state, so $\omega_{n,1}^2\geq0$. The
centrifugal term grows with $l$, so $\omega_{n,l}^2>0$ for all
$l\geq2$. In the $l=0$ sector, we find numerically exactly one negative
eigenvalue, $\omega_{0,0}^2<0$, for all $s$ considered. This is consistent with the sphaleron being a saddle point of the energy, with a single unstable direction: the growth or shrinking of the bubble.

\begin{figure}
    \centering
    \includegraphics[width=0.48\linewidth]{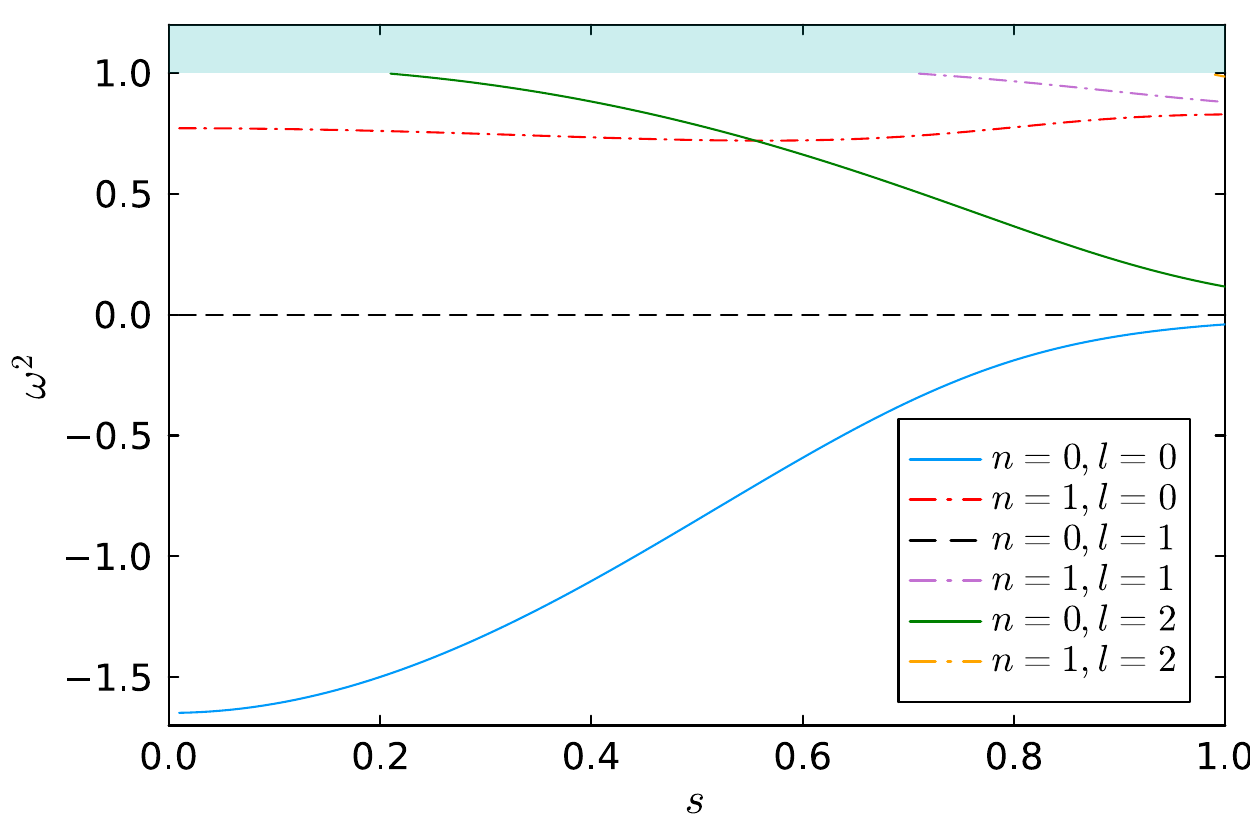}
    \includegraphics[width=0.48\linewidth]{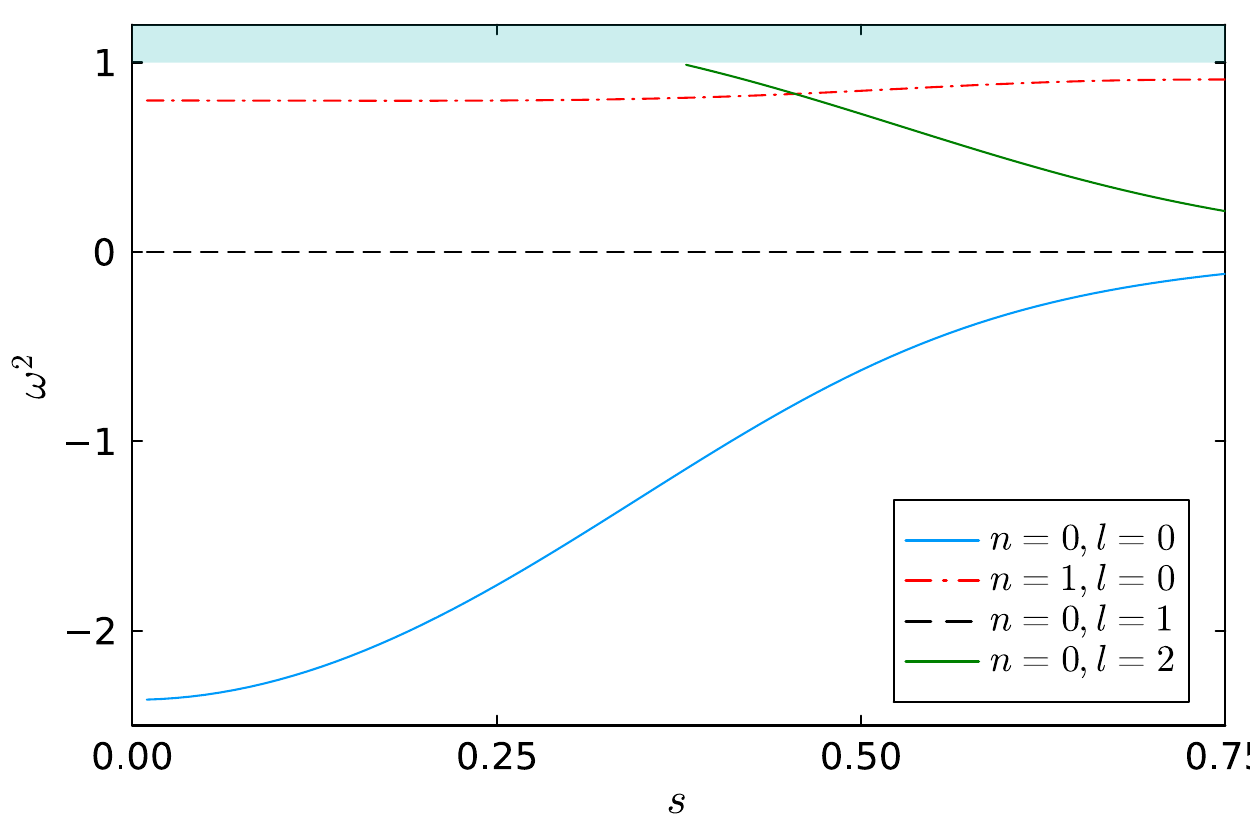}
    \caption{Lowest eigenvalues $\omega_{n,l}^2$ of the sphaleron as
    functions of $s$, for $D=2$ (left) and $D=3$ (right). The mode
    $(n,l)=(0,0)$ is unstable and $(0,1)$ is the translational zero
    mode. The shaded region, $\omega^2>1$, is the continuum.}
    \label{fig:spectrum}
\end{figure}

\begin{figure}
    \centering
    \includegraphics[width=0.48\linewidth]{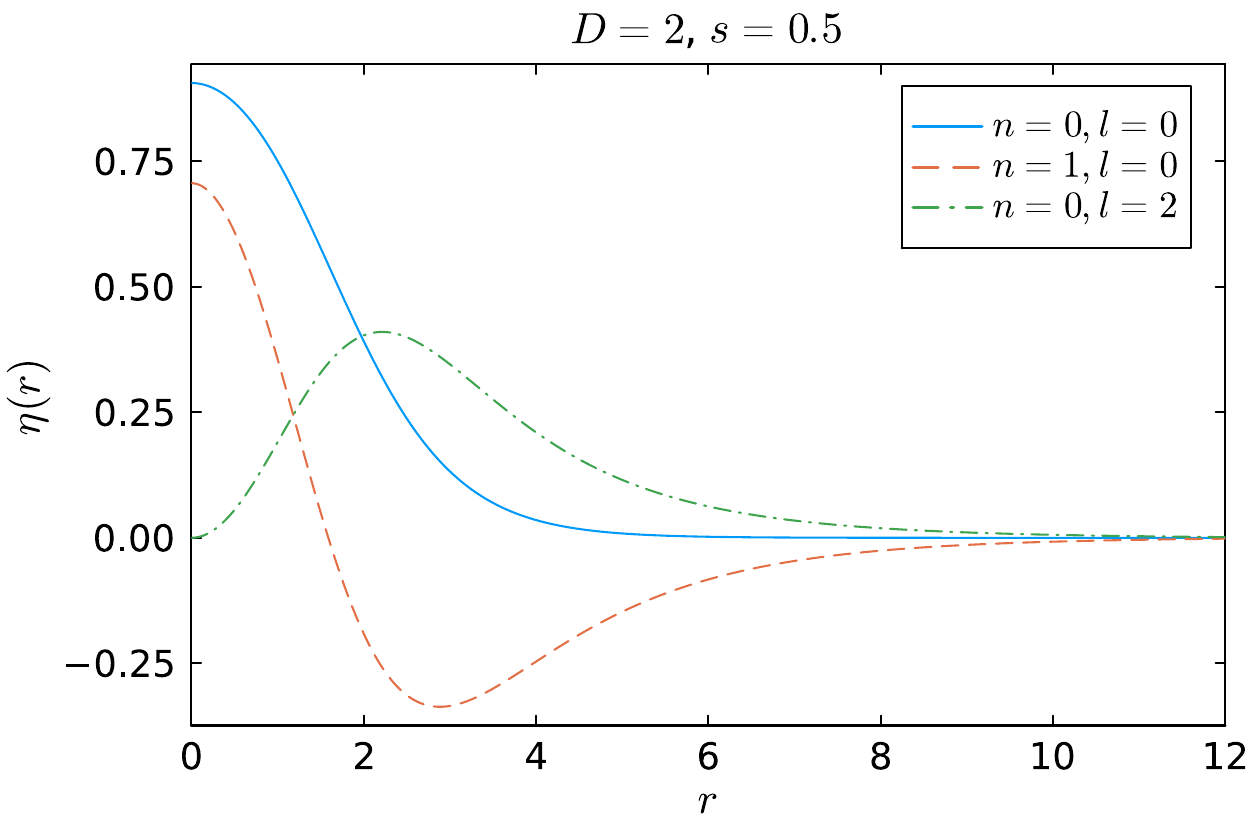}
    \includegraphics[width=0.48\linewidth]{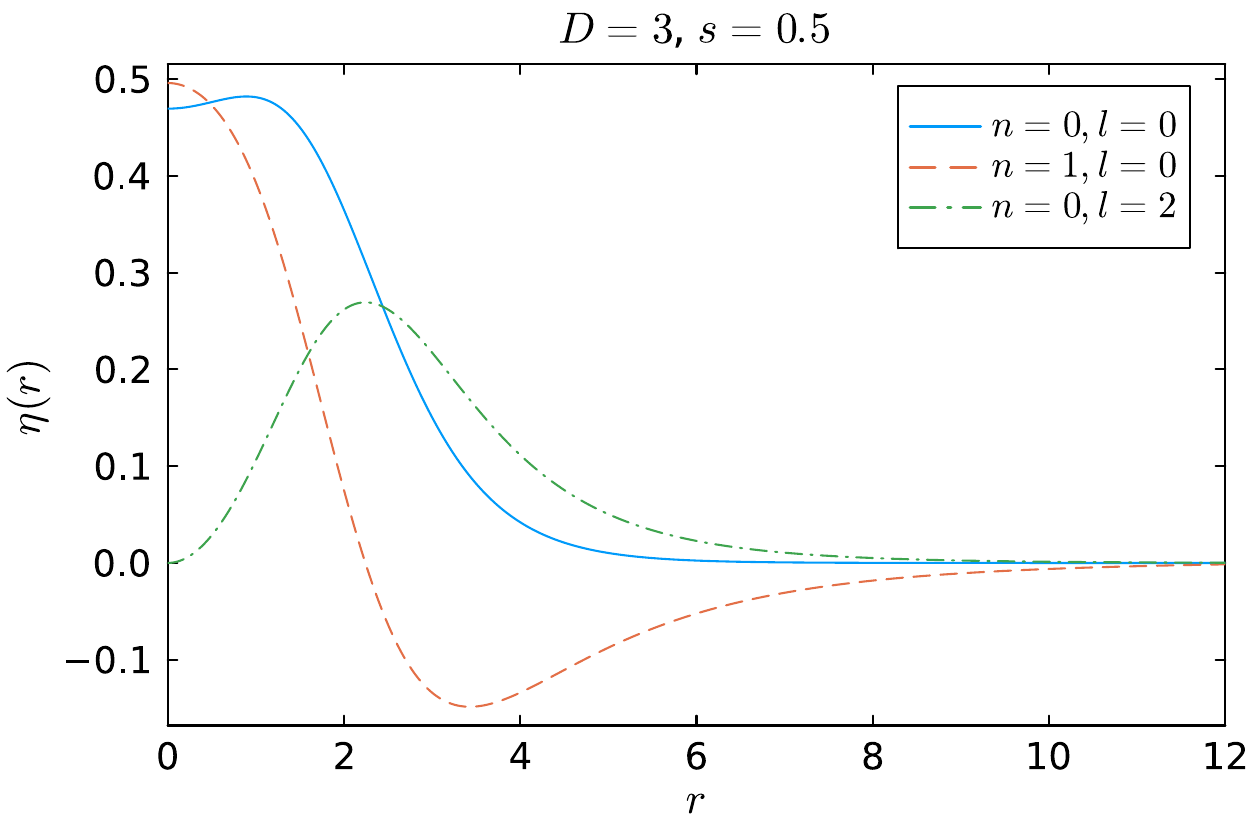}
    \caption{Unstable mode $\eta_{0,0}$, radial bound state
    $\eta_{1,0}$, and lowest nonradial bound state $\eta_{0,2}$ for
    $s=0.5$, in $D=2$ (left) and $D=3$ (right).}
    \label{fig:sphaleronustable}
\end{figure}

Figure~\ref{fig:spectrum} shows the lowest eigenvalues as functions of
$s$. As $s$ increases, $\omega_{0,0}^2$ approaches zero from below and
the instability weakens. For $s\lesssim0.38$, the instability is
stronger in $D=3$ than in $D=2$. For larger $s$, the opposite holds.
At $s=0.5$, $\omega_{0,0}^2\simeq-0.850$ ($D=2$) and $-0.625$ ($D=3$).
The radial bound state $\eta_{1,0}$ exists for all $s$ shown and is
the only one: the next radial eigenvalue already lies in the
continuum. The lowest nonradial bound state is $\eta_{0,2}$, with
$\omega_{0,2}^2\simeq0.784$ ($D=2$) and $0.729$ ($D=3$) at $s=0.5$. It
enters the continuum as $s$ decreases, at $s\approx0.2$ in $D=2$ and
$s\approx0.38$ in $D=3$. Figure~\ref{fig:sphaleronustable} shows
$\eta_{0,0}$, $\eta_{1,0}$, and $\eta_{0,2}$ at $s=0.5$. These bound
states are the localized degrees of freedom that can store energy
during the nonlinear evolution.

\section{Iterative energy exchange map}
\label{sec:map}

Resonance windows were first explained in kink-antikink collisions, as
a resonant energy exchange between the translational motion of the
kinks and their localized shape modes~\cite{campbell1983resonance}. We
formulate this mechanism as a simple iterative map, general enough to
apply to collapsing bubbles.

We describe the bounce dynamics by an effective energy $E_j$ of the
collective motion just after the $j$-th bounce, which occurs at time
$t_j$. This is not the total field energy, which is conserved, but a
measure of the distance to the boundary between escape and capture,
placed at $E_j=0$. For $E_j>0$, the system escapes, while for $E_j<0$,
it remains trapped and may undergo further bounces. For a kink-antikink pair, escape means that the kinks separate to infinity. For a collapsing bubble, it means re-expansion into the true vacuum.

Between successive bounces we write
\begin{equation}
    E_j = E_{j-1} - E_j^{\mathrm{loss}} + E_j^{\mathrm{exch}},
    \label{eq:map}
\end{equation}
where $E_j^{\mathrm{loss}}\geq 0$ is the energy lost by the collective
motion, mainly to radiation, and $E_j^{\mathrm{exch}}$ is the energy
exchanged with localized modes. Losses always push the system toward
capture. The exchange can have either sign, depending on the phase of
the localized modes at the bounce. That phase is set by the
inter-bounce time
\begin{equation}
    T_j = t_j - t_{j-1}.
    \label{eq:Tj}
\end{equation}
In Appendix~\ref{App:K-AK coll}, we show that this map reproduces the
familiar kink-antikink windows, which accumulate as a power law. In
the next subsection, we apply it to the sphaleron, where the
accumulation is instead geometric.

\subsection{Resonance mechanism of rapidly collapsing bubbles}
\label{sec:map-bubbles}

We now apply the map~\eqref{eq:map} to the decay of the sphaleron. We
perturb it along its unstable mode,
\begin{equation}
    \phi(r,0)=\phi_S(r),
    \qquad
    \dot{\phi}(r,0)=-\alpha\,\eta_{0,0}(r),
    \label{eq:initial-unstable}
\end{equation}
with $\alpha>0$ and $\eta_{0,0}>0$, so that the bubble initially
contracts. For $\int d^Dx\,\eta_{0,0}^2=1$, the initial energy above the sphaleron is
$E_{\mathrm{tot}}-E_{\mathrm{sph}}=\alpha^2/2$, which we identify with the initial effective energy $E_0$. After the first bounce, however, the effective energy
$E_j$ ($j\geq 1$) differs from this quantity, since energy is redistributed among the bubble wall, localized modes,
and radiation. Unlike the kink-antikink case, there is no obvious collective coordinate, so we use $\alpha$ as the control parameter. We denote by $\alpha_{j,c}$ the boundary, $E_j(\alpha_{j,c})=0$, between oscillon formation and decay to the true vacuum after the $j$-th bounce.

The key difference from kinks is the nature of the
boundary between the two dynamical outcomes.
A trajectory with $\alpha\simeq\alpha_{j,c}$ re-expands
after a bounce toward the sphaleron and stays near it.
Within the linear approximation, the amplitude of the
unstable mode evolves as
\begin{equation}
q(t)=q_+e^{\Gamma t}+q_-e^{-\Gamma t},
\qquad
\Gamma=\sqrt{-\omega_{0,0}^2}.
\nonumber
\end{equation}
The sign of the growing component $q_+$ determines
whether the bubble expands into the true vacuum or
collapses into an oscillon, with $q_+=0$ corresponding
to the threshold.
Assuming a smooth dependence on $\alpha$ with a
nonzero derivative at the threshold, we have, to
leading order, $q_+\propto\alpha-\alpha_{j,c}$.
Once the decaying component becomes negligible,
$q(t)\simeq q_+e^{\Gamma t}$.
We assume that the field leaves the vicinity of the
sphaleron when $|q|$ reaches a fixed value $q_*$,
independent of $\alpha$.
Setting $|q_+|e^{\Gamma T}=q_*$ then gives the time
spent near the sphaleron,
\begin{equation}
T(\alpha)\simeq\frac{1}{\Gamma}
\ln\frac{1}{|\alpha-\alpha_{j,c}|}
+\mathrm{const},
\label{eq:T-log}
\end{equation}
where the additive constant is independent of $\alpha$.
This logarithmic divergence replaces the power law
$T\propto|E|^{-1/2}$ characteristic of a weakly bound
kink-antikink pair, Eq.~\eqref{eq:T2}.

While the field stays near the sphaleron, its localized modes
oscillate with their linear frequencies. Radial dynamics excites only
$l=0$ modes, and for all parameters considered here $\eta_{1,0}$ is
the only radial bound state. Its phase when the field leaves the sphaleron is
\begin{equation}
    \Theta(\alpha)=\omega_{1,0}T(\alpha)+\delta,
    \label{eq:bubble-phase}
\end{equation}
where $\delta$ collects the phase accumulated during the nonlinear
stages. As in Appendix~\ref{App:K-AK coll}, we take
$E^{\mathrm{exch}}=\mathcal{A}\cos\Theta$ and a constant loss $L$
near threshold. Escape, $-L+\mathcal{A}\cos\Theta>0$, is possible only
if $\mathcal{A}>L$, and then occurs within windows
$|\Theta-2\pi k|<\Delta$, where $\Delta=\arccos(L/\mathcal{A})$ and
$k$ is an integer. Combining Eqs.~\eqref{eq:T-log} and
\eqref{eq:bubble-phase}, the window boundaries are
\begin{equation}
    |\alpha-\alpha_{j,c}|
    \propto
    \exp\left[-\frac{\Gamma}{\omega_{1,0}}
    \left(2\pi k\pm\Delta-\delta\right)\right].
    \label{eq:geometric}
\end{equation}
Successive windows therefore accumulate geometrically,
\begin{equation}
    \frac{|\alpha_{k+1}-\alpha_{j,c}|}{|\alpha_k-\alpha_{j,c}|}
    \simeq
    \exp\left(-\frac{2\pi\Gamma}{\omega_{1,0}}\right),
    \label{eq:geometric-ratio}
\end{equation}
and the time spent near the sphaleron increases by $2\pi/\omega_{1,0}$
from one window to the next. This is in
contrast with the power law of Eq.~\eqref{eq:powerlaw}. Using the values of Sec.~\ref{stability}, the ratio is about
$2\times10^{-4}$ for $s=0.3$ in both dimensions, and $10^{-3}$
($D=2$) and $5\times10^{-3}$ ($D=3$) for $s=0.5$, for instance. The
window pattern near $\alpha_{j,c}$ should therefore repeat in
self-similar form, each reduced by this factor. This explains why
only a few windows are visible in a standard scan.

Equations~\eqref{eq:T-log}-\eqref{eq:geometric-ratio} give three
tests: the increase of $T$ from window $k$ to $k+1$ gives
$2\pi/\omega_{1,0}$; the slope of $\ln|\alpha_k-\alpha_{j,c}|$ versus
$k$ gives $-2\pi\Gamma/\omega_{1,0}$; and, independently of the
windows, the slope of $T(\alpha)$ versus $\ln|\alpha-\alpha_{j,c}|$
gives $-1/\Gamma$. As for
kinks, radiation reduces the energy available for resonant return,
so $\mathcal{A}_j>L_j$ eventually fails at high bounce order, ending
the hierarchy of windows.

\section{Numerical results: sphaleron decay}
\label{sec:num}

\subsection{Decay along the unstable mode}
\label{sec:decay-unstable}

\begin{figure}
    \centering
    \includegraphics[width=0.9\linewidth]{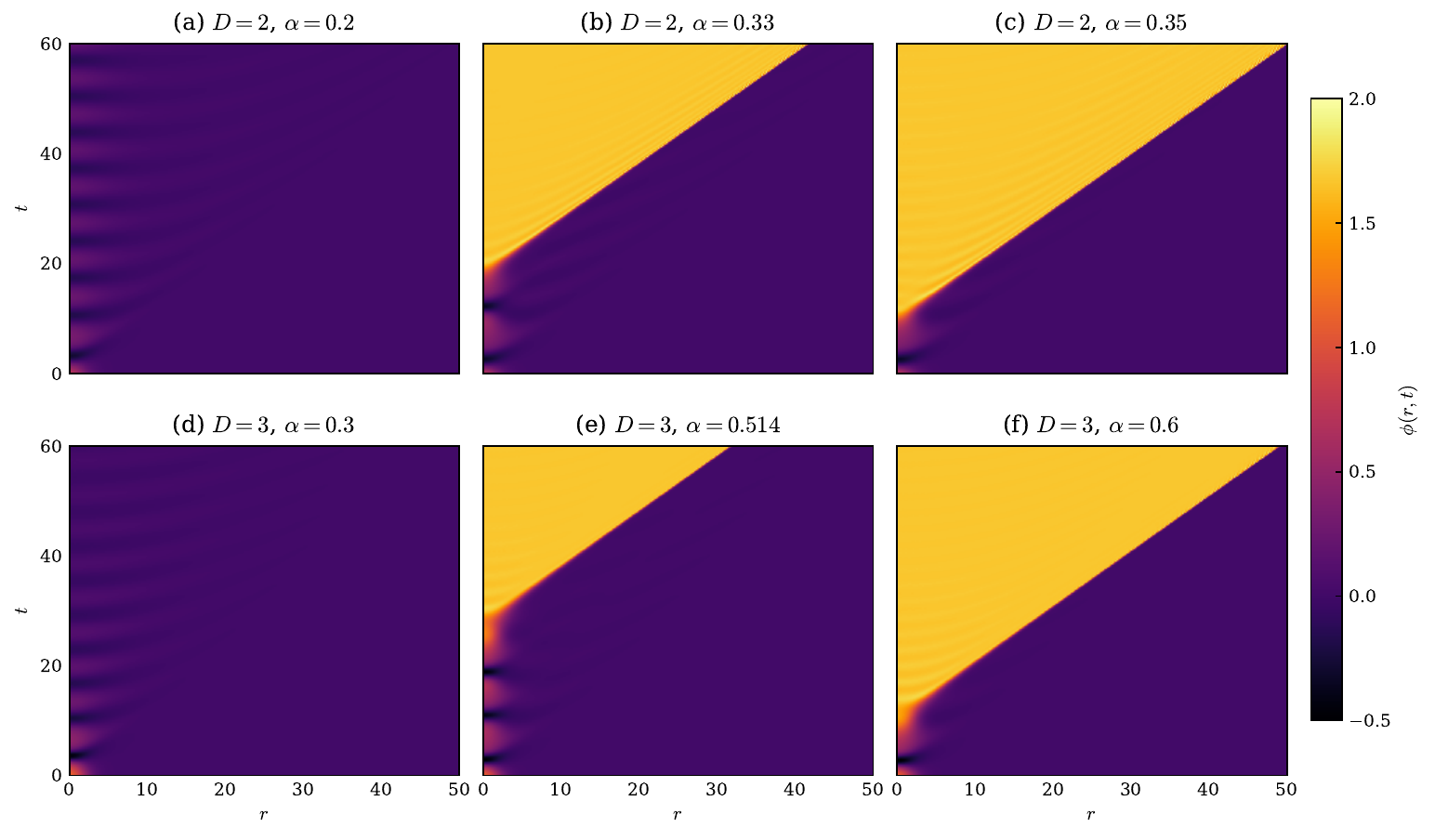}
    \caption{Field $\phi(r,t)$ for $s=0.5$ in $D=2$ (top) and $D=3$
    (bottom): oscillon formation, panels (a), (d), two- and three-bounce
    decay, panels (b), (e), and single-bounce decay, panels (c), (f).}
    \label{fig:field-vs-r-t}
\end{figure}

We evolve Eq.~\eqref{eq:eom} from the initial
conditions~\eqref{eq:initial-unstable}\footnote{A video showing the field evolution can be found on our YouTube channel. \url{https://youtu.be/asyWrkUXl6g}}. Figure~\ref{fig:field-vs-r-t}
shows the three possible outcomes for $s=0.5$. For small $\alpha$, the
collapsing bubble forms a long-lived oscillon at the origin (Figs.~\ref{fig:field-vs-r-t}(a) and \ref{fig:field-vs-r-t}(d)). For large $\alpha$, it bounces once and re-expands into
the true vacuum (Figs.~\ref{fig:field-vs-r-t}(c) and \ref{fig:field-vs-r-t}(f)). In between it can bounce
several times before decaying to the true vacuum (Figs.~\ref{fig:field-vs-r-t}(b) and \ref{fig:field-vs-r-t}(e)).
Although only the unstable mode is excited initially, these
multi-bounce solutions indicate that energy is temporarily stored in
other degrees of freedom, as described in Sec.~\ref{sec:map}.

\begin{figure}
    \centering
    \includegraphics[width=0.85\linewidth]{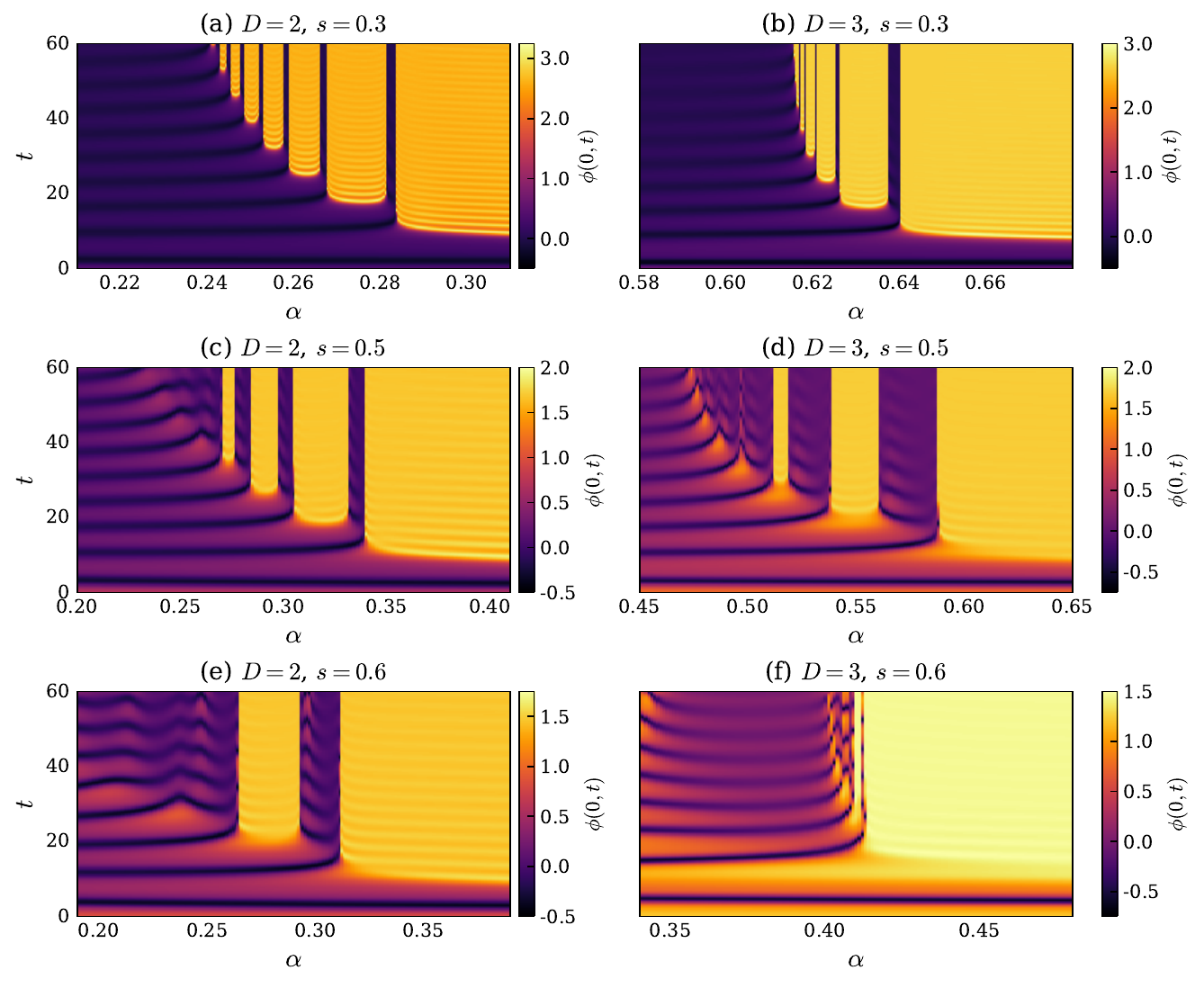}
    \caption{Field at the origin, $\phi(0,t)$, as a function of
    $\alpha$ and $t$ for $D=2$ (left) and $D=3$ (right), with
    $s=0.3$, $0.5$, and $0.6$ (top to bottom).}
    \label{fig:field-vs-alpha-t}
\end{figure}

Figure~\ref{fig:field-vs-alpha-t} shows the field at the origin,
$\phi(0,t)$, as a function of $\alpha$ and $t$, for several values of
$s$ in $D=2$ and $D=3$. Oscillations around $\phi=0$ indicate an
oscillon, while a plateau at $\phi_+$ indicates decay to the true
vacuum. The two regimes are separated by the one-bounce critical value
$\alpha_c$. Below $\alpha_c$, resonance windows, in which the bubble
decays after two or more bounces, alternate with regions of oscillon
formation. The number of bounces increases by one from window to
window as $\alpha$ decreases. As $s$ increases, the windows become
wider and fewer. 

\begin{figure}
    \centering
    \includegraphics[width=0.48\linewidth]{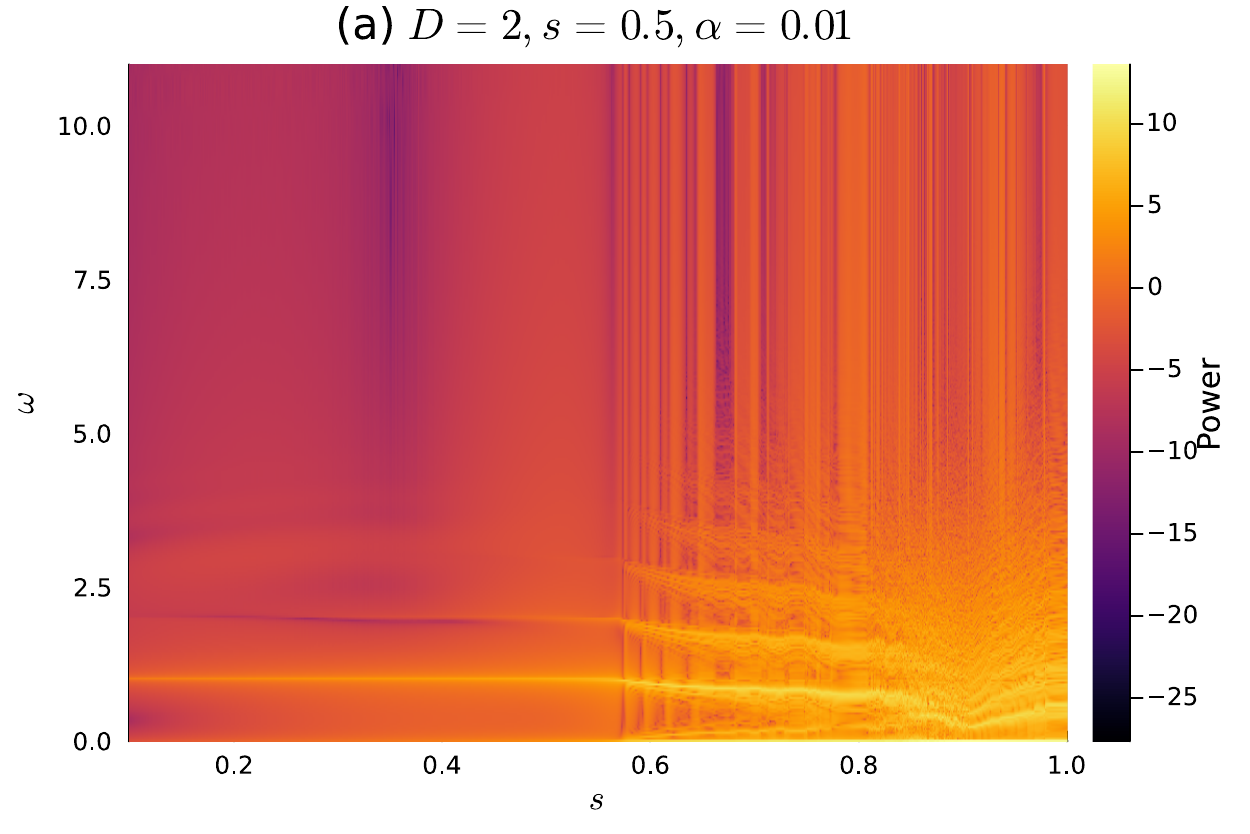}
    \includegraphics[width=0.48\linewidth]{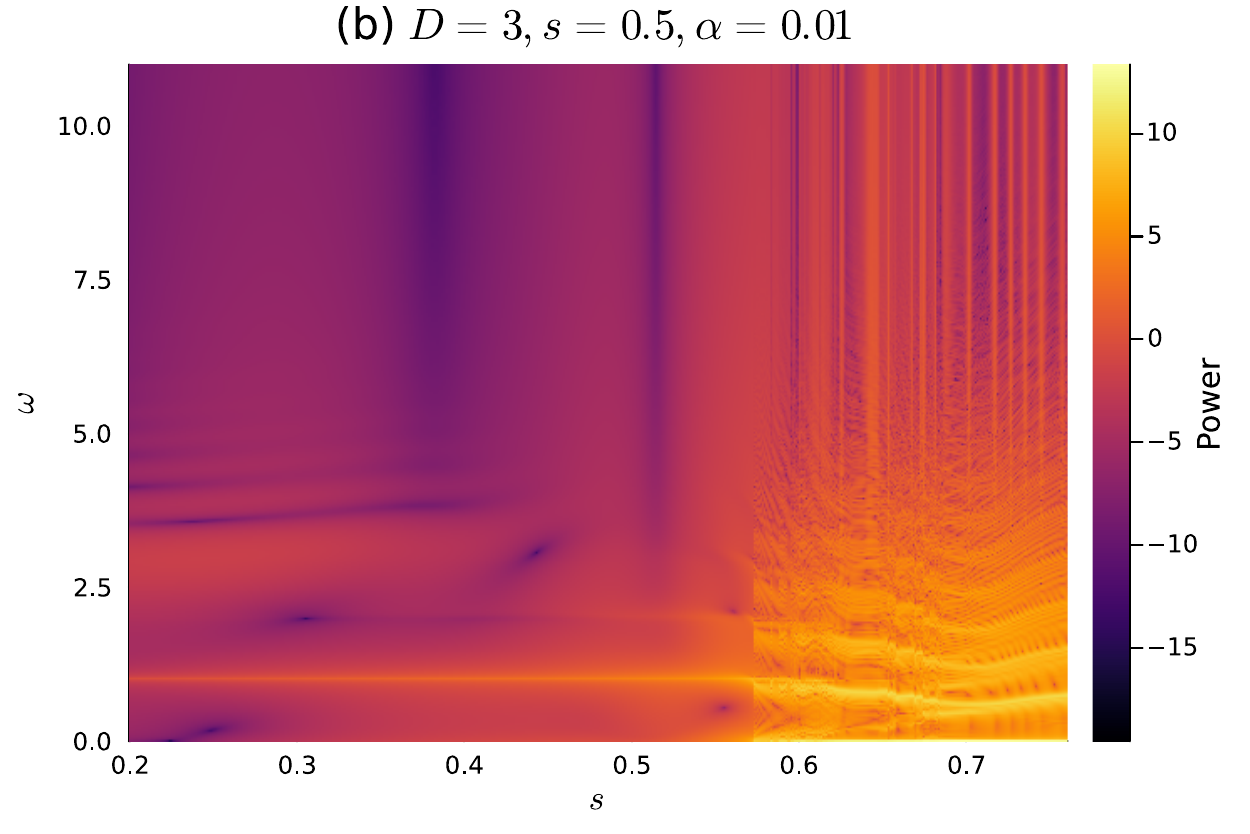}
    \caption{Power spectrum (logarithmic scale) of $\phi(0,t)$ for the
    oscillon formed with $\alpha=0.01$, as a function of $s$, for
    $D=2$ (left) and $D=3$ (right).}
    \label{fig:Fourier_spectrum}
\end{figure}

To characterize the oscillon, Fig.~\ref{fig:Fourier_spectrum} shows
the spectrum of $\phi(0,t)$, sampled with $\Delta t=0.2$ over
$N=1000$ steps after the initial transient. For small $s$, the
spectrum is that of a small-amplitude oscillon: a frequency just below
the mass threshold $\omega=1$ and its harmonics~\cite{fodor2008small}.
For larger $s$, the sphaleron and the resulting oscillon are larger,
and additional frequencies appear. This coincides with the reduced
number of windows in Figs.~\ref{fig:field-vs-alpha-t}(e) and (f). It
suggests that energy spread over more oscillon modes is less
efficiently returned to the bubble wall.

\begin{table}[t]
    \centering
    \begin{tabular}{l|cc|cc}
        $D$ & \multicolumn{2}{c|}{2} & \multicolumn{2}{c}{3} \\
        $s$ & 0.3 & 0.5 & 0.3 & 0.5 \\
        \hline
        $\omega_{1,0}$          & 0.865 & 0.850 & 0.896 & 0.922 \\
        $\omega_{\mathrm{fit}}$ & 0.859 & 0.876 & 0.895 & 0.964 \\
        \hline
        $\Gamma/\omega_{1,0}$   & 1.332 & 1.085 & 1.381 & 0.857 \\
        $(\Gamma/\omega_{1,0})_{\mathrm{fit}}$
                                & 1.270 & 1.046 & 1.306 & 0.864 \\
        \hline
        $\Gamma$                & 1.151 & 0.922 & 1.237 & 0.790 \\
        $\Gamma_{\mathrm{fit}}$ & 1.090 & 0.916 & 1.168 & 0.833 \\
    \end{tabular}
    \caption{Linear frequency $\omega_{1,0}$ of the radial bound state
    and growth rate $\Gamma=\sqrt{-\omega_{0,0}^2}$ of the unstable
    mode, compared with the values fitted near $\alpha_c$:
    $\omega_{\mathrm{fit}}$ from the decay times of successive windows,
    Eq.~\eqref{eq:bubble-phase}; $(\Gamma/\omega_{1,0})_{\mathrm{fit}}$
    from the window positions, Eq.~\eqref{eq:geometric}; and
    $\Gamma_{\mathrm{fit}}$ from the time spent near the sphaleron,
    Eq.~\eqref{eq:T-log}.}
    \label{tab:frequencies}
\end{table}

\begin{figure}
    \centering
    \includegraphics[width=0.9\linewidth]{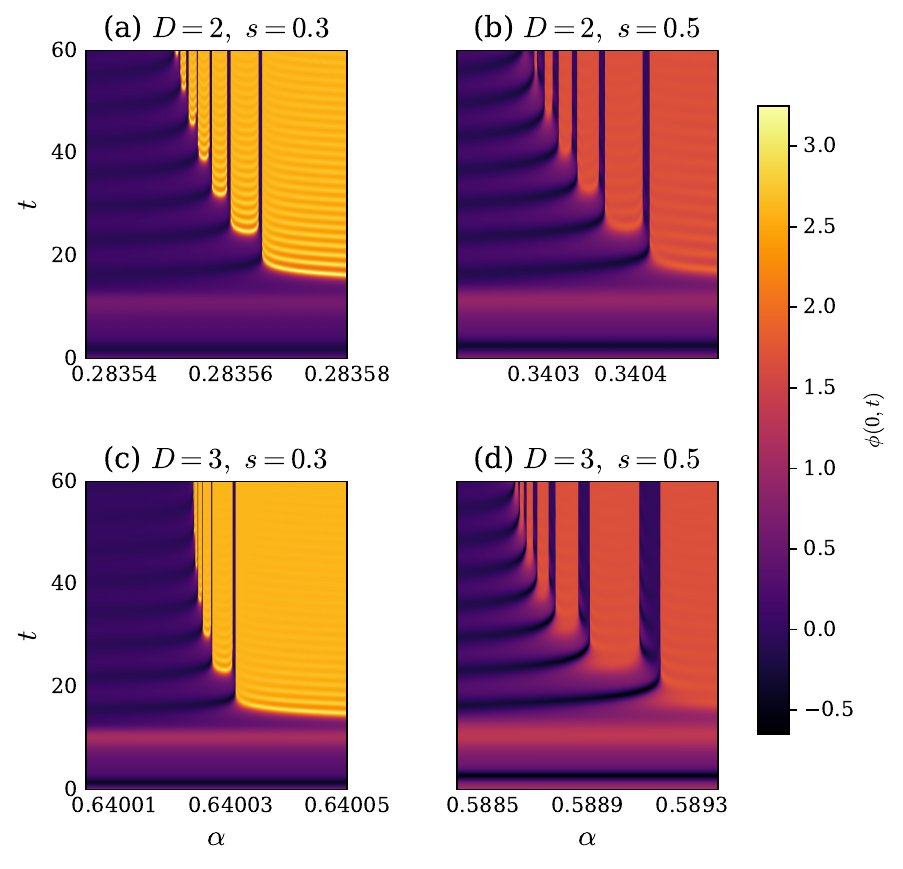}
    \caption{Zoom of Fig.~\ref{fig:field-vs-alpha-t} near the
    one-bounce threshold $\alpha_c$, for $D=2,3$ and $s=0.3,0.5$.}
    \label{fig:zoom}
\end{figure}

\begin{figure}
    \centering
    \includegraphics[width=0.9\linewidth]{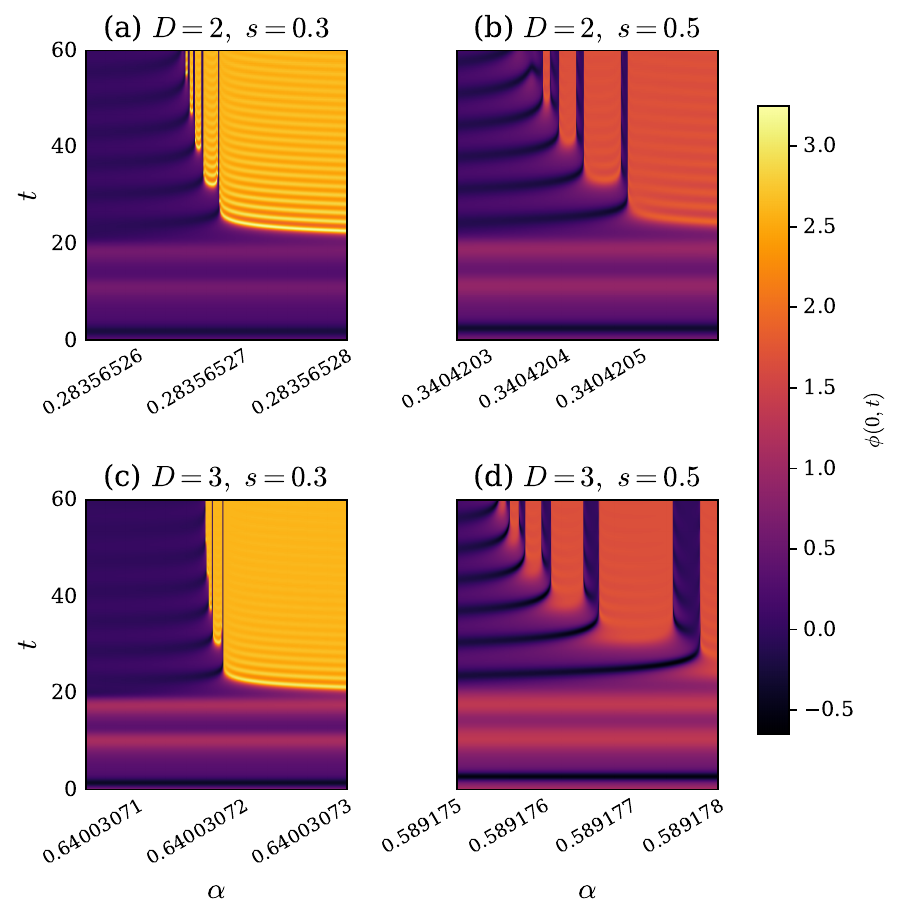}
    \caption{Zoom of Fig.~\ref{fig:zoom} near the one-bounce threshold $\alpha_c$ for $D=2,3$ and $s=0.3,0.5$.}
    \label{fig:zoom2}
\end{figure}

The geometric accumulation predicted in Sec.~\ref{sec:map-bubbles}
implies that most windows near $\alpha_c$ are too narrow to be seen in
Fig.~\ref{fig:field-vs-alpha-t}. Zooming in by several orders of
magnitude, Fig.~\ref{fig:zoom} shows a new set of multi-bounce
windows that repeats the structure of Fig.~\ref{fig:field-vs-alpha-t}
on a much smaller scale in $\alpha$. To perform the tests of
Sec.~\ref{sec:map-bubbles}, we located three successive two-bounce
windows, $k=1,2,3$: the first is visible in
Fig.~\ref{fig:field-vs-alpha-t}, the second in Fig.~\ref{fig:zoom}
and the third only at a further zoom in Fig.~\ref{fig:zoom2} . The results are listed in Table~\ref{tab:frequencies}. All fitted values agree with
the linear spectrum of the sphaleron within $6\%$, supporting the
logarithmic law~\eqref{eq:T-log} and the role of the radial bound
state $\eta_{1,0}$ in the resonant energy exchange.

\subsection{Excitation of the radial bound state}
\label{sec:exc-sph-10}

To understand the role of the radial bound state $\eta_{1,0}$ directly, we
excite it together with the unstable mode,
\begin{equation}
    \phi(r,0)=\phi_S(r),
    \qquad
    \dot\phi(r,0)=-\alpha\,\eta_{0,0}(r)+\beta\,\eta_{1,0}(r),
    \label{IC-phid10}
\end{equation}
with $\eta_{1,0}(0)>0$. The bound state thus starts with zero
displacement, and $\beta$ is its initial velocity amplitude. For normalized modes, the initial energy exceeds the sphaleron
energy by $(\alpha^2+\beta^2)/2$, independently of the sign of $\beta$.

\begin{figure}
    \centering
    \includegraphics[width=0.95\linewidth]{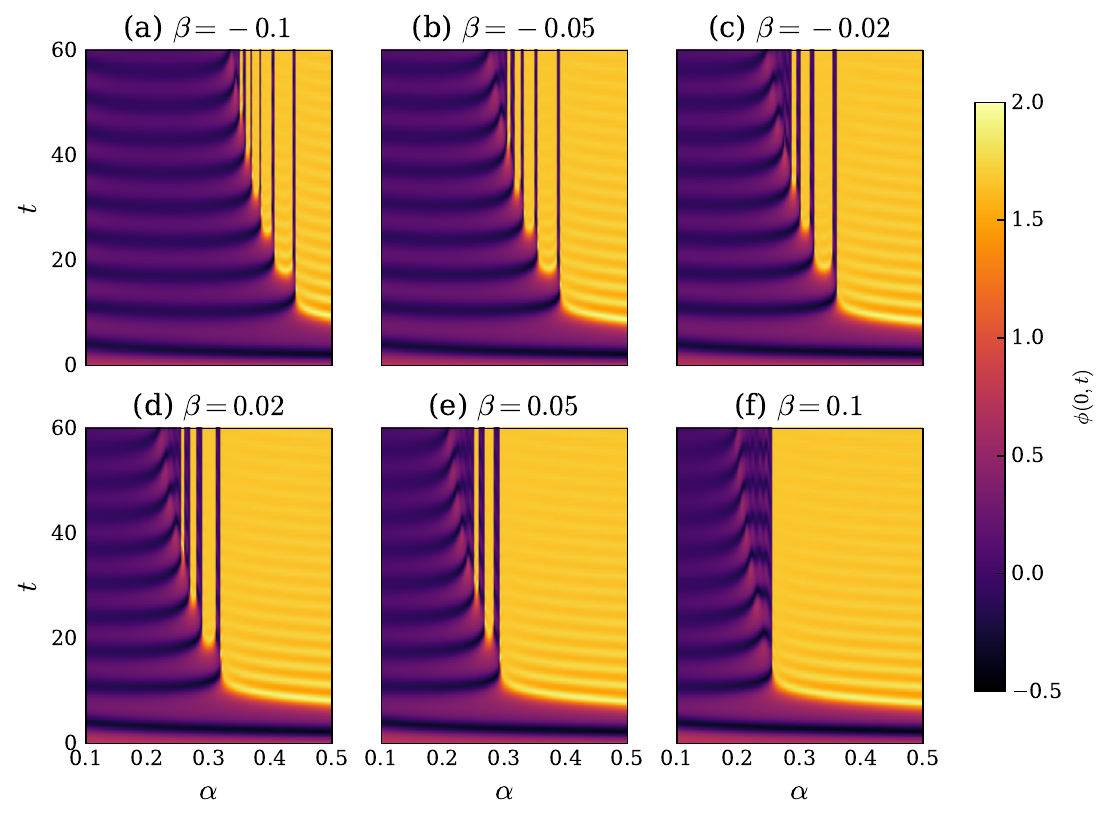}
    \caption{Field at the origin, $\phi(0,t)$, as a function of
    $\alpha$ and $t$ for $D=2$ and $s=0.5$, with the radial bound
    state $\eta_{1,0}$ excited with amplitude $\beta$,
    Eq.~\eqref{IC-phid10}.}
    \label{fig:field-vs-alpha-t-2D-betal0}
\end{figure}

\begin{figure}
    \centering
    \includegraphics[width=0.95\linewidth]{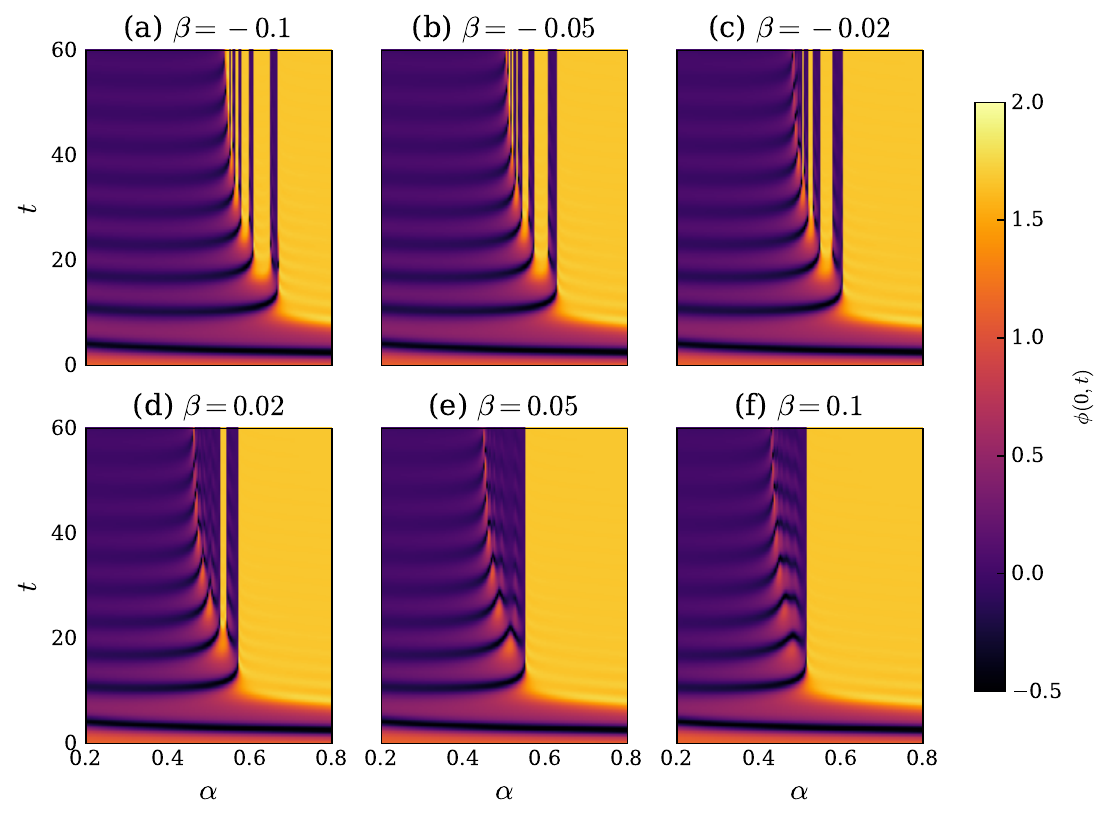}
    \caption{Same as Fig.~\ref{fig:field-vs-alpha-t-2D-betal0}, for
    $D=3$.}
    \label{fig:field-vs-alpha-t-3D-betal0}
\end{figure}

Figures~\ref{fig:field-vs-alpha-t-2D-betal0} and
\ref{fig:field-vs-alpha-t-3D-betal0} show $\phi(0,t)$ as a function
of $\alpha$ for several values of $\beta$, with $s=0.5$. Even a small
excitation of the bound state changes the window structure significantly,
both in the position and in the number of windows. 

The dynamics also
depends on the sign of $\beta$: the threshold $\alpha_c$ decreases for
$\beta>0$ and increases for $\beta<0$.
This sign dependence follows from the initial motion of the bubble
wall. We define the bubble radius $R(t)$ by
$\phi(R(t),t)=\phi_{\mathrm{w}}$, with $\phi_{\mathrm{w}}=\phi_-$, the
top of the potential barrier. Differentiating in time gives
$\dot R=-\partial_t\phi/\partial_r\phi$ at $r=R$, so that
\begin{equation}
    \dot R(0)
    =\frac{\alpha\,\eta_{0,0}(R_0)-\beta\,\eta_{1,0}(R_0)}
          {\phi_S'(R_0)},
    \label{eq:wall-velocity}
\end{equation}
where $\phi_S(R_0)=\phi_{\mathrm{w}}$. For $s=0.5$, $R_0\simeq2.1$
($D=2$) and $2.7$ ($D=3$), beyond the node of $\eta_{1,0}$, so
$\eta_{1,0}(R_0)<0$. Since $\phi_S'(R_0)<0$, positive $\beta$
enhances the initial contraction of the bubble, while negative $\beta$
opposes it. This conclusion holds for any $\phi_{\mathrm{w}}$ below
the value of $\phi_S$ at the node, about $0.44$ ($D=2$) and $0.47$
($D=3$). For a fixed $\alpha$ and $|\beta|$, the total initial
energy is independent of the sign of $\beta$. However, positive $\beta$ enhances the initial contraction of the bubble wall, favoring vacuum
decay at smaller $\alpha$, while negative $\beta$ has the opposite effect.

\subsection{Quadrupolar excitation in $D=2$}
\label{sec:exc-sph-02-2D}

We now break spherical symmetry by exciting the lowest nonradial
vibrational mode, $\eta_{0,2}$, with angular dependence
$\cos 2\theta=(x^2-y^2)/r^2$. The initial conditions are
\begin{align}
\phi(\mathbf{x},0)&=\phi_S(r)+\beta\,\eta_{0,2}(r)\cos2\theta,  \nonumber\\
    \dot\phi(\mathbf{x},0)&=-\alpha\,\eta_{0,0}(r)
    +\beta\,\omega_{0,2}\,\eta_{0,2}(r)\cos2\theta,
    \label{IC-02-2D}
\end{align}
so that, at linear order, the mode oscillates as
$\sqrt2\,\beta\,\eta_{0,2}\cos2\theta\,\cos(\omega_{0,2}t-\pi/4)$.

Two symmetry properties constrain the dynamics. First, the
transformation $\beta\to-\beta$ is equivalent to a rotation by
$\pi/2$, since $\cos[2(\theta+\pi/2)]=-\cos2\theta$. Therefore, the
evolution at the origin is exactly independent of the sign of $\beta$.
Second, regularity requires $\eta_{0,2}\sim r^2$ near the origin, so $\eta_{0,2}(0)=0$
and the quadrupolar mode does not change $\phi(0,t)$ at linear order. Moreover, since
$\cos 2\theta$ averages to zero over angles, it does not contribute to the spherically
symmetric ($l=0$) part of the field, which is responsible for the collapse of the bubble. The quadrupolar mode therefore affects $\phi(0,t)$ and the collapse only through nonlinear
coupling. Its lowest-order effect is of order $\beta^2$, through the $l=0$ component of $\cos^2 2\theta = (1+\cos 4\theta)/2$.

\begin{figure}
    \centering
    \includegraphics[width=0.95\linewidth]{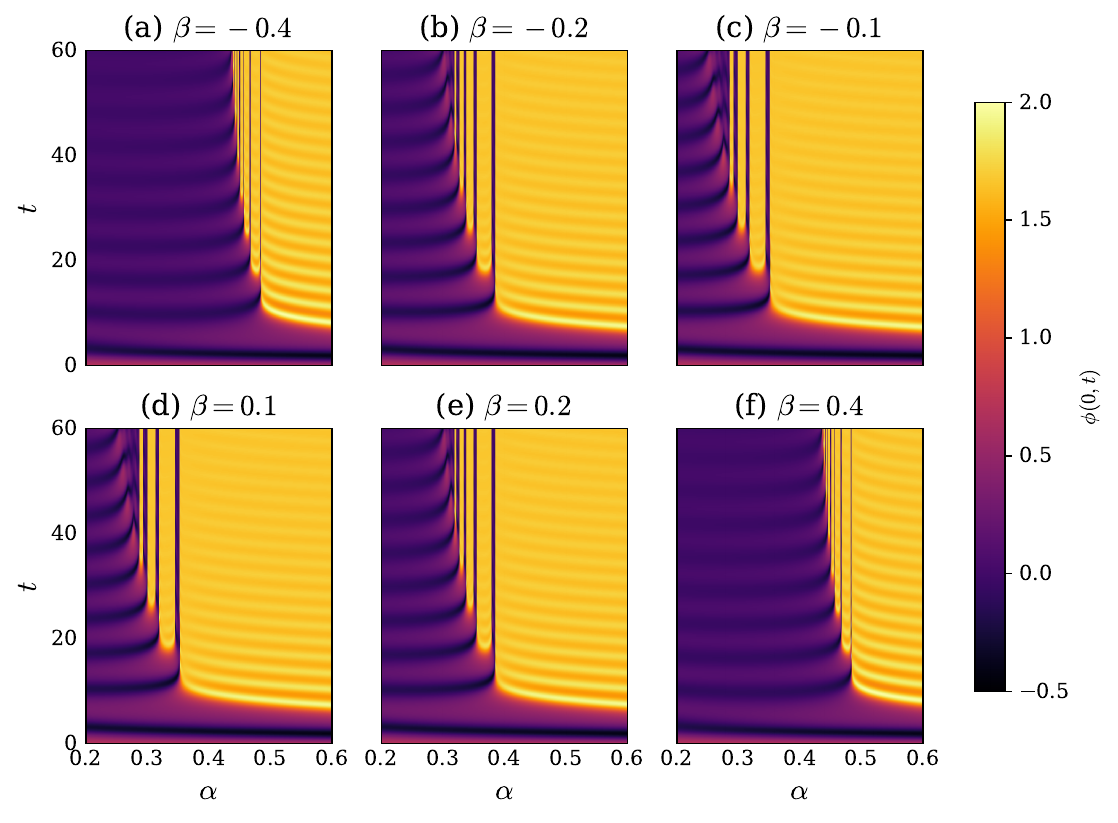}
    \caption{Field at the origin, $\phi(0,t)$, as a function of
    $\alpha$ and $t$ for $D=2$ and $s=0.5$, with the quadrupolar mode
    $\eta_{0,2}$ excited with amplitude $\beta$, Eq.~\eqref{IC-02-2D}.}
    \label{fig:field-vs-alpha-t-2D-betal2}
\end{figure}

Figure~\ref{fig:field-vs-alpha-t-2D-betal2} shows $\phi(0,t)$ as a
function of $\alpha$ for several values of $\beta$, with $s=0.5$.
Panels with opposite $\beta$ coincide, as required by symmetry. For
small $|\beta|$, the structure is close to the purely radial case of
Fig.~\ref{fig:field-vs-alpha-t}. As $|\beta|$ increases, the threshold
$\alpha_c$ moves to larger values: part of the energy is stored in the
nonradial deformation, so a stronger radial perturbation is required
for decay to the true vacuum.

\subsection{Quadrupolar excitation in $D=3$}
\label{sec:exc-sph-02-3D}

In three dimensions, we excite the axisymmetric quadrupolar mode,
\begin{align}
    \phi(\mathbf{x},0)&=\phi_S(r)+\beta\,\eta_{0,2}(r)\,Y_2^0(\theta),
    \nonumber\\
    \dot\phi(\mathbf{x},0)&=-\alpha\,\eta_{0,0}(r)
    +\beta\,\omega_{0,2}\,\eta_{0,2}(r)\,Y_2^0(\theta),
    \label{IC-02-3D}
\end{align}
with $Y_2^0(\theta)=\sqrt{5/16\pi}\,(3\cos^2\theta-1)$, using the same
convention as in Eq.~\eqref{IC-02-2D}. 

Since $Y_2^0>0$ at the poles and $Y_2^0<0$ at the equator, $\beta>0$
produces a prolate bubble and $\beta<0$ an oblate one. Unlike in
$D=2$, these two configurations cannot be related by a rotation, so
the dynamics may depend on the sign of $\beta$.

The form of this dependence can be understood from a perturbative
expansion in $\beta$. At linear order, the quadrupolar excitation
does not affect the spherically symmetric component of the field,
since its spherical average vanishes,
$\langle Y_2^0\rangle=\frac{1}{4\pi}\int Y_2^0\,d\Omega=0$. However,
nonlinear mode coupling can generate spherically symmetric
contributions at quadratic and cubic order, since
$\langle(Y_2^0)^2\rangle\neq0$ and $\langle(Y_2^0)^3\rangle\neq0$.
Assuming a smooth dependence of the threshold on $\beta$, we
therefore expect
\begin{equation}
    \alpha_c(\beta)-\alpha_c(0)
    =c_2\beta^2+c_3\beta^3+\mathcal{O}(\beta^4).
    \label{eq:alphac-beta}
\end{equation}
The cubic term gives the leading difference between prolate and
oblate bubbles. In $D=2$, by contrast, rotational symmetry requires
$\alpha_c(\beta)=\alpha_c(-\beta)$, so only even powers of $\beta$
appear.

\begin{figure}
    \centering
    \includegraphics[width=0.95\linewidth]{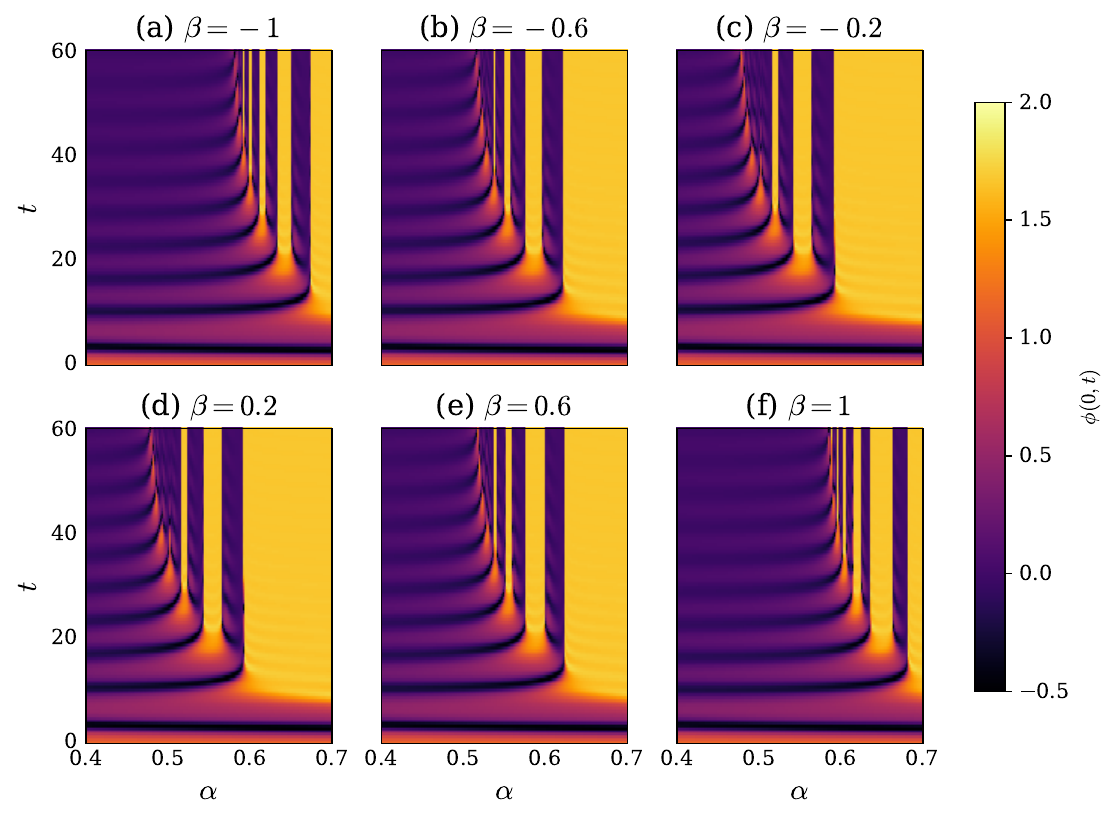}
    \caption{Field at the origin, $\phi(0,t)$, as a function of
    $\alpha$ and $t$ for $D=3$ and $s=0.5$, with the quadrupolar mode
    $\eta_{0,2}Y_2^0$ excited with amplitude $\beta$,
    Eq.~\eqref{IC-02-3D}.}
    \label{fig:field-vs-alpha-t-3D-betal2}
\end{figure}

Figure~\ref{fig:field-vs-alpha-t-3D-betal2} shows $\phi(0,t)$ as a
function of $\alpha$ for $\beta=\pm0.2$, $\pm0.6$, and $\pm1$, with
$s=0.5$. Oscillons and single-bounce decay are again separated by
resonance windows, and $\alpha_c$ increases with $|\beta|$. We find
$\alpha_c=0.5923$, $0.6237$, and $0.6823$ for $\beta=0.2$, $0.6$,
and $1$, and $\alpha_c=0.5923$, $0.6217$, and $0.6737$ for the
corresponding negative values. These values agree with
Eq.~\eqref{eq:alphac-beta}. The even part grows approximately as
$\beta^2$, while the odd part,
$[\alpha_c(\beta)-\alpha_c(-\beta)]/2$, grows as $\beta^3$: it equals
$0.0010$ at $\beta=0.6$ and $0.0043$ at $\beta=1$, and is negligible
at $\beta=0.2$. The prolate deformation thus raises the threshold for vacuum decay slightly more than
the oblate one. The resonance windows shift by a comparable amount, but they are
noticeably wider for prolate than for oblate bubbles.

\section{Conclusion}
\label{sec:conclusion}

We have studied the decay of spherical sphalerons, the critical bubbles
of a scalar field theory with a false vacuum, in $D=2$ and $D=3$ spatial dimensions. The collapse is initiated by a velocity perturbation along the unstable mode with amplitude $\alpha$. For small
$\alpha$, the bubble collapses into a long-lived oscillon. Above a critical value $\alpha_c$, it bounces once and expands indefinitely, triggering false-vacuum decay. In between, resonance windows
appear in which the bubble bounces several times before expanding. Thus, a perturbation that drives the bubble toward collapse, rather than expansion, can still trigger false-vacuum decay.

We have shown that the resonance windows near $\alpha_c$ accumulate geometrically, with a scaling ratio $e^{-2\pi\Gamma/\omega_{1,0}}$ determined by the linear spectrum of the sphaleron. We explain this behavior with an effective energy-exchange map. Near the threshold, the
field remains close to the sphaleron for a time that diverges as
$\Gamma^{-1}\ln(1/|\alpha-\alpha_c|)$. During this time, the radial bound state oscillates with frequency $\omega_{1,0}$, and its phase determines whether energy is returned to the radial motion, allowing
the bubble to expand.
This mechanism differs from that of conventional kink-antikink collisions, where the time between successive collisions diverges as a power law,
leading to a power-law accumulation of resonance windows. The geometric accumulation gives rise to a self-similar
window structure, with successive scales differing by several orders of magnitude.
This explains why only a few windows are found in standard numerical scans.
The values extracted from the nonlinear dynamics agree with the predictions based on the linear spectrum within $6\%$, supporting the proposed mechanism.

We have also studied how initially exciting the bound states affects the decay. Exciting the radial bound state $\eta_{1,0}$ significantly modifies the window structure and shifts the threshold $\alpha_c$ in a direction set by the bound state's contribution to the bubble wall's initial velocity. The quadrupolar bound state $\eta_{0,2}$ couples to the radial collapse only nonlinearly and raises the threshold. In $D=2$, invariance under rotation by $\pi/2$ makes the shift even in the quadrupolar amplitude. In $D=3$, a cubic term is
allowed, and prolate bubbles have a slightly higher threshold than
oblate ones.

The mechanism described here involves two key ingredients: an unstable mode that determines the time spent near the sphaleron and a localized bound state that allows resonant energy exchange. We expect a similar geometric accumulation of resonance windows in other systems where trajectories
near the threshold between two dynamical outcomes spend a long time close to a static solution with a single unstable mode and a bound state that couples to the collective motion. It would be interesting to investigate whether this mechanism also occurs in the thin-wall regime and whether it plays a role in the evolution of bubbles
nucleated by thermal or quantum fluctuations during first-order phase transitions in the early Universe.


\appendix

\section{Kink-antikink collisions}
\label{App:K-AK coll}

Let us apply the map~\eqref{eq:map} to the collision of a kink and
an antikink of mass $M$ and velocities $\pm v$. Here the boundary is
the escape of the pair to infinity, so the initial effective energy is
the nonrelativistic kinetic energy, $E_0=Mv^2$. The one-bounce
critical velocity $v_{1,c}$ is defined by $E_1(v_{1,c})=0$: for
$v>v_{1,c}$ the pair escapes after the first collision, while for $v$
slightly below $v_{1,c}$ it is weakly bound and collides again. Near
threshold,
\begin{equation}
    E_1(v)\simeq b_1\left(v-v_{1,c}\right),\qquad b_1>0.
    \label{eq:E1}
\end{equation}
Following Campbell \textit{et al.}~\cite{campbell1983resonance}, the
time between the first and second collisions is
\begin{equation}
    T_2(v)\simeq\frac{a_2}{\sqrt{v_{1,c}^2-v^2}}\propto |E_1|^{-1/2},
    \label{eq:T2}
\end{equation}
where $a_2$ is a constant. It diverges as $v\to v_{1,c}^-$.

Suppose the first collision excites a shape mode of frequency
$\omega_S$. At the second collision its phase is
\begin{equation}
    \Theta_2(v)=\omega_S T_2(v)+\delta_2,
    \label{eq:phase2}
\end{equation}
where $\delta_2$ is a phase shift accumulated during the collisions.
We model the phase-dependent exchange and the loss at the second
collision as
\begin{equation}
E_2^{\mathrm{exch}}=\mathcal{A}_2\cos\Theta_2,\qquad E_2^{\mathrm{loss}}\simeq L_2,
    \qquad \mathcal{A}_2,L_2>0.
    \label{eq:redist2}
\end{equation}
Since $E_1\to0$ at threshold, Eq.~\eqref{eq:map} gives \footnote{$E_1$
is negligible in the energy balance, but not in $T_2$, which diverges
as $E_1\to0$.}
\begin{equation}
    E_2(v)\simeq -L_2+\mathcal{A}_2\cos\Theta_2(v).
    \label{eq:E2}
\end{equation}
The pair escapes after the second collision if $E_2>0$, which requires
$\mathcal{A}_2>L_2$: the mode must return more energy than is lost.
Defining $\Delta_2=\arccos(L_2/\mathcal{A}_2)\in(0,\pi/2)$, the escape
condition reads $|\Theta_2-2\pi k|<\Delta_2$, with $k$ an integer such
that $T_2>0$. Using Eqs.~\eqref{eq:T2} and \eqref{eq:phase2}, the
$k$-th two-bounce window is $v_{2,k}^-<v<v_{2,k}^+$, with
\begin{equation}
    \left(v_{2,k}^{\pm}\right)^2
    = v_{1,c}^2
    -\frac{a_2^2\omega_S^2}{\left(2\pi k\pm\Delta_2-\delta_2\right)^2}.
    \label{eq:windows}
\end{equation}
For large $k$, Eq.~\eqref{eq:windows} gives
\begin{equation}
    v_{1,c}^2-\left(v_{2,k}^{\pm}\right)^2\propto k^{-2},
    \qquad
    \left(v_{2,k}^{+}\right)^2-\left(v_{2,k}^{-}\right)^2\propto k^{-3}.
    \label{eq:powerlaw}
\end{equation}
The first relation gives the distance of the $k$-th window from the
critical velocity, and the second its width. Although every window
has the same width $2\Delta_2$ in phase, $\Theta_2$ varies ever faster
near $v_{1,c}$, so the windows become both closer to $v_{1,c}$ and
narrower. The two-bounce windows thus accumulate at $v_{1,c}$ as a
power law, the familiar structure of kink-antikink
collisions~\cite{campbell1983resonance}.
This follows from the
power-law divergence of $T_2$ in Eq.~\eqref{eq:T2}.

Higher-bounce windows follow in the same way. Near the edge of a $(j-1)$-bounce
window, only the last inter-bounce time $T_j$ diverges, and the earlier ones can be
absorbed into the phase shift, $\Theta_j = \omega_S T_j + \delta_j$. Then
$E_j \simeq -L_j + \mathcal{A}_j \cos\Theta_j$, and a $j$-bounce window requires
$\mathcal{A}_j > L_j$. Each window edge therefore hosts a new accumulation of
windows, which produces the nested structure~\cite{campbell1986kink}. Since
radiation reduces the energy available for resonant return, this condition
eventually fails, ending the observable hierarchy of windows.

\section{Numerical Methods}

Both the sphaleron profiles and the linear perturbation spectrum are obtained using the shooting method, with $\phi_S(0)$ and $\omega^2$ as the shooting parameters, respectively. In both cases, we use a regular power-series expansion near the origin to circumvent the coordinate
singularity and integrate outward using a fourth-order Runge-Kutta method. The shooting parameters are determined
by bisection.

The time evolution of the radial equation of motion is obtained as follows. The radial coordinate is discretized with a spacing of $\Delta r=0.05$, and the outer boundary is set at $L=100.0$. Spatial derivatives are computed using a five-point stencil approximation. The resulting equations of motion are then integrated in time using the Tsit5 integrator from DifferentialEquations.jl. Near the boundaries, we lower the order of the finite difference approximation for the derivatives. Moreover, at $r=L$, we consider the Dirichlet boundary condition $\phi(r=L)=0$. To guarantee regularity at the origin, the field is assumed to be an even function. We also consider that $\phi'/r\to\phi''$ as $r\to 0$.

The time evolution of the complete equations of motion, used for the
quadrupolar excitations of Secs.~\ref{sec:exc-sph-02-2D} and
\ref{sec:exc-sph-02-3D}, is obtained as follows. In both cases, we use the
symmetry of the initial data to reduce the computational domain. For $D=2$,
the initial conditions~\eqref{IC-02-2D} are even in $x$ and in $y$, a
property preserved by Eq.~\eqref{eq:eom}, so we evolve only the quadrant
$0\leq x,y\leq L$, with $L=40$ and $\Delta x=\Delta y=0.05$. For $D=3$, the
initial conditions~\eqref{IC-02-3D} do not depend on the azimuthal angle and
are even in $z$. The problem then reduces to the quadrant $0\leq\rho,z\leq L$
in cylindrical coordinates, with $L=60$ and $\Delta\rho=\Delta z=0.15$, where
$\nabla^2=\partial_\rho^2+\rho^{-1}\partial_\rho+\partial_z^2$. On the axis,
we use $\rho^{-1}\partial_\rho\phi\to\partial_\rho^2\phi$. Spatial derivatives
are computed using a five-point stencil approximation. The profiles $\phi_S$, $\eta_{0,0}$ and $\eta_{0,2}$ are interpolated from the radial solutions onto the grid. The equations are integrated in time with a fixed-step fourth-order Runge-Kutta method, with $\Delta t=0.02$ ($D=2$) and $\Delta t=0.03$ ($D=3$), using
Numba-compiled Python code. The boxes are large enough that radiation
reflected at the boundaries does not reach the origin. For
$\beta=0$, the axisymmetric code reproduces the evolution of the radial
equation.
\footnote{The source codes used to generate the figures presented in this work are available at
\url{https://github.com/joao-gfc/Numerical-Simulations-of-Oscillons}.}


\section*{Acknowledgements}
AM acknowledges financial support from CNPq (Conselho Nacional de
Desenvolvimento Científico e Tecnológico), Grant No.~306295/2023-7,
and from CAPES (Coordenação de Aperfeiçoamento de Pessoal de Nível
Superior). FCS acknowledges financial support from the
PQ-C-12556/25 Produtividade Estaduais CNPq/FAPEMA program, CNPq Grant No. 402830/2023-7 and No. 420896/2025-2 and Coordenação de Aperfeiçoamento de Pessoal de Nível Superior (CAPES) - Finance Code 001 (Brazilian agencies).


\bibliographystyle{unsrt}
\bibliography{references}

\end{document}